\documentclass[12pt]{article}
\usepackage{fullpage}
\usepackage{graphicx}
\newcommand{\be}{\begin{equation}}
\newcommand{\ee}{\end{equation}}

\newcommand{\cartoon}{{\em cartoon\ }}

\newcommand{\news}{\setcounter{equation}{0}\quad}
\def\ben{\begin{equation}}
\def\een{\end{equation}}
\def\bea{\begin{eqnarray}}
\def\eea{\end{eqnarray}}
\begin{document}
\title{
\begin{flushright}\ \vskip -2cm {\normalsize{\em DCPT-08/69}}\end{flushright}\vskip 1cm
\Large {\bf VORTON CONSTRUCTION AND DYNAMICS}}
\author{
Richard A. Battye$^{1}$
and Paul M. Sutcliffe$^{2}$\\[10pt]
\\{\normalsize $^{1}$
{\sl Jodrell Bank Centre for Astrophysics,} }
\\{\normalsize {\sl University of Manchester,
 Manchester M13 9PL, U.K.}}
\\{\normalsize {\sl Email : Richard.Battye@manchester.ac.uk}}\\
\\{\normalsize $^{2}$
{\sl \normalsize Department of Mathematical Sciences,
Durham University, Durham DH1 3LE, U.K.}}\\
{\normalsize {\sl Email: p.m.sutcliffe@durham.ac.uk}}\\[10pt]}
\date{December 2008}
\maketitle
\begin{abstract}
Vortons are closed loops of superconducting cosmic strings
carrying current and charge. In this paper we present the
first numerical construction of vortons in
the global version of Witten's $U(1)\times U(1)$ theory.
An energy minimization procedure is used to compute stationary
vortons for a range of charges and currents, and the
associated vorton radius is calculated.
It is found that the standard analysis based on infinite straight
cosmic strings does not provide a good description
of the vorton cross-section.
The computed solutions are used as initial conditions in an
axially symmetric time evolution code, which verifies that the 
solutions are indeed stationary, and are stable to axially
symmetric perturbations. 
Perturbations which preserve the axial symmetry excite 
oscillatory modes and produce an evolution which eventually 
returns to a stationary vorton.
Finally, the constructed vorton solutions are used 
as initial conditions in a full (3+1)-dimensional simulation
and an instability to non-axial perturbations is found.
The instability produces a pinching and bending of the vorton
which results in its destruction. 
\end{abstract}

\newpage
\section{Introduction}\news
Cosmic strings are topological defects which may have formed during
a phase transition in the early universe (for a review see
\cite{VS}). Superconducting cosmic strings were introduced by
 Witten \cite{Wi1} and possess a non-dissipative current flowing
along the string due to the coupling of a second complex scalar field.
A closed loop of superconducting string, carrying both current and charge,
is known as a vorton \cite{DS2}. It is a stationary solution
in which the current and charge on the string provide a
force to balance the string tension and prevent its collapse.

Vortons have a number of potential cosmological consequences \cite{BCDT} 
and it has been suggested that they may be involved in cosmological phenomena
that include galactic magnetic
fields, high energy cosmic rays, gamma ray bursts and baryogenesis.
It is therefore of considerable interest to determine the 
stationary properties and dynamical behaviour of vortons.

Despite a number of studies, the existence and stability of vortons as 
classical field theory solutions is still an open problem, with
little convincing evidence, even numerically. Numerical simulations of the
relevant nonlinear field theory are difficult to perform in
(3+1)-dimensions and results are limited, mainly due to the
existence of multiple length and time scales. The problem is
also compounded
by a large parameter space in which to search. The simplest model
expected to have vorton solutions is the global version of Witten's 
$U(1)\times U(1)$ theory \cite{Wi1}, but even in this theory vorton
solutions have not been found. 
The only field theory computation to date \cite{LS} is
in a modified version of this theory, in which the
interaction term between the two complex scalar fields is replaced
by a non-renormalizable interaction, designed to make the numerical
problem more tractable. In this modified model a vorton 
has been presented \cite{LS}, although even this eventually decays. 

In this paper we address the aspects discussed above,
by presenting the first numerical construction of vortons in
the global version of Witten's $U(1)\times U(1)$ theory with renormalizable interactions.
An energy minimization procedure is used to compute stationary
vortons for a range of charges and currents, and the
associated vorton radius is calculated.
The computed solutions are then used as initial conditions in an
axially symmetric time evolution code and this verifies that the 
solutions are indeed stationary. 
It is shown that even small axial perturbations of these 
initial conditions excite reasonably large amplitude long-lived oscillations,
though the evolution does produce a slow relaxation back to a 
stationary vorton.
Considerable computational resources are applied to remove the
above axially symmetric constraint and perform fully (3+1)-dimensional 
simulations.
These results appear to show that the constructed solutions have an
instability to non-axial perturbations. 
The instability produces a pinching and bending of the vorton
which results in its destruction.  

Vortons have been analysed in a thin string limit by approximating
the loop cross-section by that of an infinite straight
cosmic string carrying current and charge \cite{LS}.
In a planar analogue, known as kinky vortons \cite{BS}, a similar
analysis was found to be in excellent agreement with full numerical
simulations of the nonlinear field theory. However, we find that
the straight string analysis does not provide a good description 
of the vortons presented in this paper, and we suggest some reasons
for this.
Finally, we discuss the relevance of parameter choices for vorton
existence and study how the solutions vary as parameters are changed.

\section{The model and parameters}\news
The Lagrangian density for the global version of  Witten's 
$U(1)\times U(1)$ model \cite{Wi1}  is given by
\be
{\cal L}=\partial_\mu\phi \partial^\mu \bar\phi
+\partial_\mu \sigma \partial^\mu \bar\sigma
-\frac{\lambda_\phi}{4}(|\phi|^2-\eta_\phi^2)^2
-\frac{\lambda_\sigma}{4}(|\sigma|^2-\eta_\sigma^2)^2
-\beta|\phi|^2 |\sigma|^2+\frac{\lambda_\sigma}{4}\eta_\sigma^4,
\label{lag}
\ee
where $\phi$ and $\sigma$ are complex scalar fields,
with $\eta_\phi,\eta_\sigma,\lambda_\phi,\lambda_\sigma,\beta$ all 
real positive constants. 

The theory has a global $U(1)\times U(1)$ symmetry and the parameters
can be arranged so that in the vacuum the $U(1)$ symmetry 
associated with $\phi$ is
broken, $|\phi|=\eta_\phi\ne 0,$ while the $U(1)$ symmetry 
associated with $\sigma$ remains unbroken, $\sigma=0.$
This requires that 
\be
\lambda_\phi \eta_\phi^4>\lambda_\sigma \eta_\sigma^4.
\label{convac}
\ee

For this symmetry breaking pattern there exist global vortex strings 
constructed from the $\phi$ field and the interaction term makes it
possible that the $\sigma$ field can form a condensate inside the 
string core.

One of the difficulties in studying vortons is that the parameter space
of the theory (\ref{lag}) is quite large and it is not clear which
parameter values are compatible with the existence of vortons. 
Furthermore, one requires vortons that can be computed numerically with
reasonable computing resources. This is a significant issue because
typically one expects a range of length and time scales to be present
in the problem and an accurate numerical resolution could easily require
unacceptable resources.

Kinky vortons \cite{BS} are (2+1)-dimensional analogues of vortons,
in which the role of the cosmic string is replaced by a kink string.
In fact the Lagrangian density is precisely the (2+1)-dimensional
version of (\ref{lag}) with the restriction that $\phi$ is real.
In two space dimensions numerical computations require less resources
and also some aspects are more analytically tractable. Studies of
kinky vortons suggest that the parameter set
\be \eta_\phi=1, \quad
\eta_\sigma=1, \quad \lambda_\phi=3, \quad
\lambda_\sigma=2, \quad \beta=2 \label{param3}, \ee
may be favourable for the existence of vortons with scales
that are feasible for numerical investigation. 
From now on we restrict to this parameter set, though in 
Section~\ref{sec-inf} we shall discuss results for other parameter choices.

A stationary vorton is a solution of the time dependent field equations
that follow from (\ref{lag}). The $\phi$ field is static and axially
symmetric, describing a circular global string loop. The time 
dependence of the $\sigma$ field is in the form of a phase rotation with
constant frequency, which induces a non-zero value of the Noether
charge $Q$ associated with the unbroken global symmetry 
\be
Q=\frac{1}{2i}\int (\dot\sigma\bar\sigma-\dot{\bar\sigma}\sigma)\, d^3x.
\label{charge}
\ee
Not only does the vorton carry charge $Q$ but it also carries a current,
associated with a phase winding of the $\sigma$ field along 
the loop. There is an associated integer $N,$ usually referred to as the 
winding number, which counts the number of total twists of the phase of 
$\sigma$ in a complete circuit of the loop. Note that although $N$ is
traditionally called the winding number and is integer-valued, it is not
a topological quantity. It becomes ill-defined if $\sigma$ vanishes
at any point on the circuit around which the winding is defined, and hence
can change with time during a continuous time evolution. 

String tension will favour a reduction in the radius of a vorton loop
but the expectation is that the charge and current in the $\sigma$ field
can provide a force balance that results in a preferred finite non-zero
radius. Of course, this will only be the case if the $\sigma$ condensate 
remains localized around the string core, and this is far from guaranteed
as there is at most a finite energy barrier to be overcome for the two
to separate. Analytical and numerical investigations \cite{LS} make 
it clear that a delicate balance between charge $Q$ and winding $N$ 
(or current) is required, and certainly if either $Q$ or $N$ 
vanish then there is no vorton solution.
  
The task is to find suitable values for the quantities $Q$ and 
$N,$ for which a vorton exists and can be computed, and hence 
determine its properties, in particular the loop radius.
As a vorton solution is not static this can not be achieved using
a simple energy minimization approach, but a modification that
includes charge conservation can be applied. This is discussed 
in the following Section and results are presented.    

\section{Vorton construction via energy minimization}\news
For fields that have a time dependence of the form
 $\sigma=e^{i\omega t}\psi,$ where $\psi$ and $\phi$ are independent
of time,  the energy can be written as 
\bea
E&=&\int \left\{\partial_i\phi \partial_i \bar\phi
+\partial_i\psi \partial_i \bar\psi
+\frac{\lambda_\phi}{4}(|\phi|^2-\eta_\phi^2)^2
+\frac{\lambda_\sigma}{4}(|\psi|^2-\eta_\sigma^2)^2
+\beta|\phi|^2 |\psi|^2-\frac{\lambda_\sigma}{4}\eta_\sigma^4
\right\}d^3x\cr
&+&\frac{Q^2}{\int |\psi|^2 \, d^3x},
\label{energy}
\eea
where we have used the relation
\be
Q=\omega{\int |\psi|^2 \, d^3x}.
\label{omega}
\ee

The fields of a stationary vorton take the axially 
symmetric form
\be
\phi(x,y,z)=\phi(\rho,0,z), \quad\quad
\psi(x,y,z)=e^{iN\theta}\psi(\rho,0,z),
\label{axial}
\ee
where $\rho$ and $\theta$ are polar coordinates in the $(x,y)$ plane,
so that $\theta=0$ corresponds to $(x,y)=(\rho,0).$
Without loss of generality, the field $\psi(\rho,0,z)$ can be
chosen to be real.

The fields $\phi(\rho,0,z)$ and $\psi(\rho,0,z)$ are determined
by minimization of the energy (\ref{energy}), using a gradient
flow algorithm. Rather than solving the problem numerically using
cylindrical polar coordinates, the \cartoon
 method \cite{cartoon} is employed. This involves using Cartesian
coordinates, but restricting the simulations to the slice $y=0.$ \, 
Derivatives are 
approximated using a second order accurate finite difference scheme 
with a lattice spacing $\Delta x.$ 
 To implement the finite difference
approximations the fields are also required on the slices $(x,\pm \Delta x,z),$
but these are obtained using the axially symmetric ansatz (\ref{axial})
by mapping back to the slice $y=0,$ and interpolating from the
grid in this slice. 
The \cartoon method has been successfully used to compute other examples
of axially symmetric solitons \cite{IS}.

\begin{figure}[ht]
\begin{center}
\includegraphics[width=14cm]{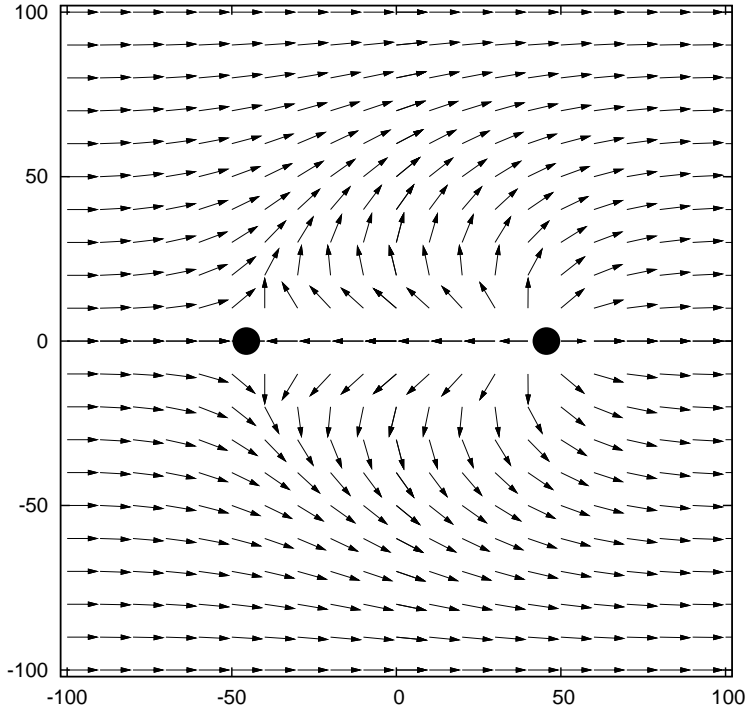}
\caption{The field $\phi(x,0,z)$ for a vorton with $Q=13000$
and $N=25,$ giving a radius $R=45.6.$ 
The horizontal and vertical length of each arrow is 
proportional to the real and imaginary parts of $\phi.$ The black dots
denote the points at which $\phi$ vanishes. 
}
\label{fig-arrows}
\end{center}
\end{figure}

In addition to the axial symmetry (\ref{axial}) there is also a reflection
symmetry 
\be
\phi(x,y,-z)=\bar\phi(x,y,z), \quad  \psi(x,y,-z)=\psi(x,y,z),
\label{relfection}
\ee
therefore the computation can be reduced to the quarter plane $x \ge 0$
and $z\ge 0.$ The boundary conditions along the symmetry axis $\rho=0$
are $\partial_\rho \phi=0$ and $\psi=0.$ Considering the whole plane
$y=0,$ then the vorton configuration has the form of a vortex anti-vortex pair
with positions on the positive and negative $x$-axis.
Along a contour at infinity in this plane the winding number of $\phi$ is zero,
 and therefore the fields take the constant vacuum values
 $\phi=\eta_\phi$  and $\psi=0$ along this contour. 
On a finite lattice the computational range is limited to
$-L\le x \le L$ and $-L\le z \le L$ and the fields are set to the
vacuum values on this computational boundary.  
Most of the simulations presented here use a lattice spacing $\Delta x=0.5$
and have the boundary at $L=100$ or $L=200,$ depending upon the 
size of the vorton to be computed.

Figure~\ref{fig-arrows} displays the field $\phi(x,0,z),$ 
for a vorton solution with  $Q=13000$ and $N=25.$  
The horizontal and vertical components of each arrow are
proportional to the real and imaginary parts of $\phi,$
and this displays the winding properties of the field. The field is
constant along the boundary, and hence has zero winding number in the
plane.
Along the $z$-axis the field winds exactly once, corresponding to the
fact that the winding in the right half plane $z\ge 0$ is $+1$
and in the left half plane $z\le 0$ is $-1.$ It is important to note
that the field is far from the vacuum value over a $z$ range that is 
of the order of the vorton radius, therefore a thin numerical
grid (with less points in the $z$ direction) can not be used, even
though this may appear appropriate at first glance.

\begin{figure}[ht]
\begin{center}
\includegraphics[width=12cm]{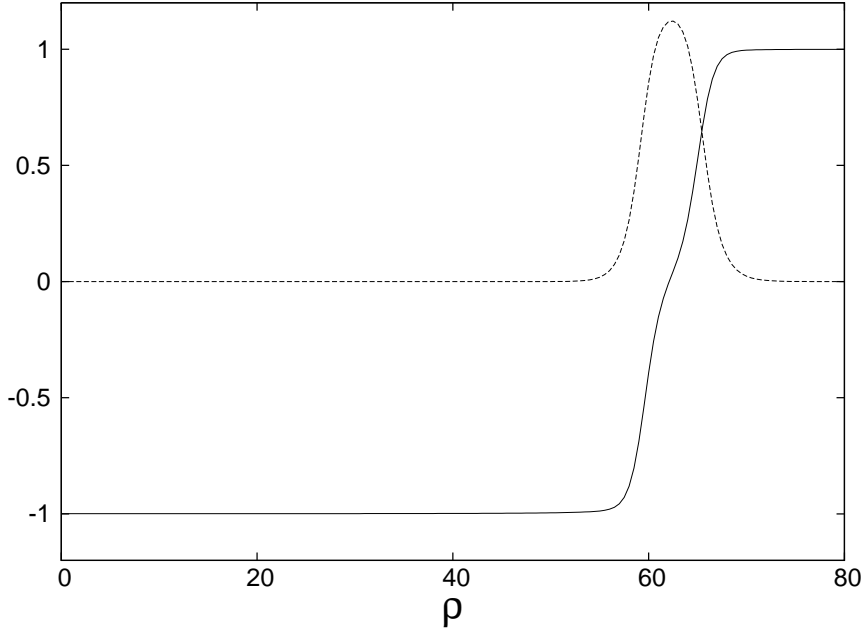}
\caption{The fields $\Re(\phi)$ (solid curve) and $|\sigma|$ 
(dashed curve) as a function 
of $\rho$ in the plane $z=0$, for a vorton with $Q=13000$ and $N=45,$
corresponding to $R=62.6.$}
\label{fig-N45}
\end{center}
\end{figure}

\begin{figure}[ht]
\begin{center}
\includegraphics[width=12cm]{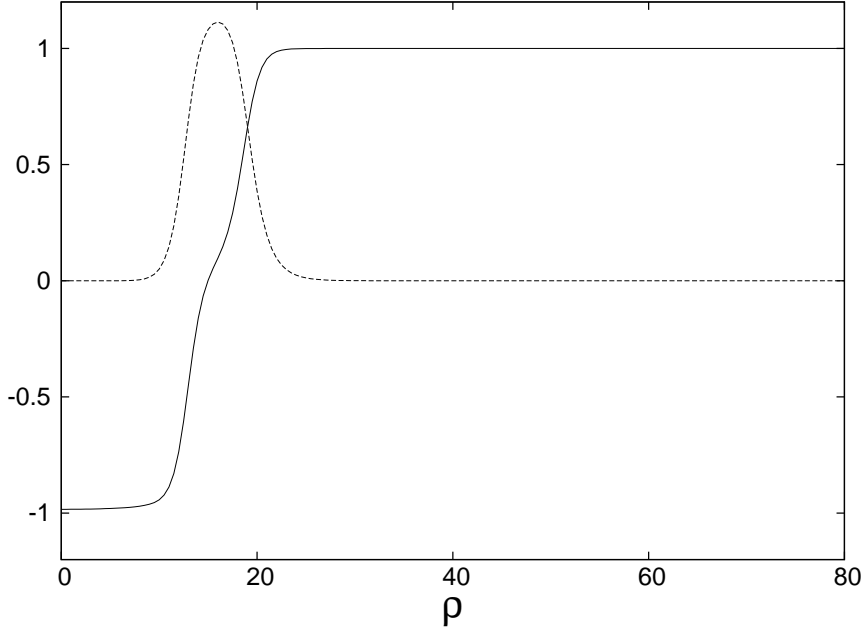}\caption{
The fields $\Re(\phi)$ (solid curve) and $|\sigma|$ 
(dashed curve) as a function 
of $\rho$ in the plane $z=0$, for a vorton with $Q=3000$ and $N=10,$
corresponding to $R=15.5.$
}
\label{fig-N10}
\end{center}
\end{figure}

Along the positive $x$-axis $\phi$ is real and has a kink-like form, with
the vorton radius $R$ defined by $\phi(R,0,0)=0.$
The values of $\phi$ and $|\sigma|$ along this axis are
plotted in Figure~\ref{fig-N45},
for the vorton with $Q=13000$ and $N=45,$
corresponding to $R=62.6,$ and in Figure~\ref{fig-N10}
for the vorton with $Q=3000$ and $N=10,$ corresponding to $R=15.5.$
Notice that even though the first of these vortons is much
larger than the second, the profiles are similar after a translation.
This aspect is discussed further in Section~\ref{sec-inf}.

\begin{figure}[ht]
\begin{center}
\includegraphics[width=12cm]{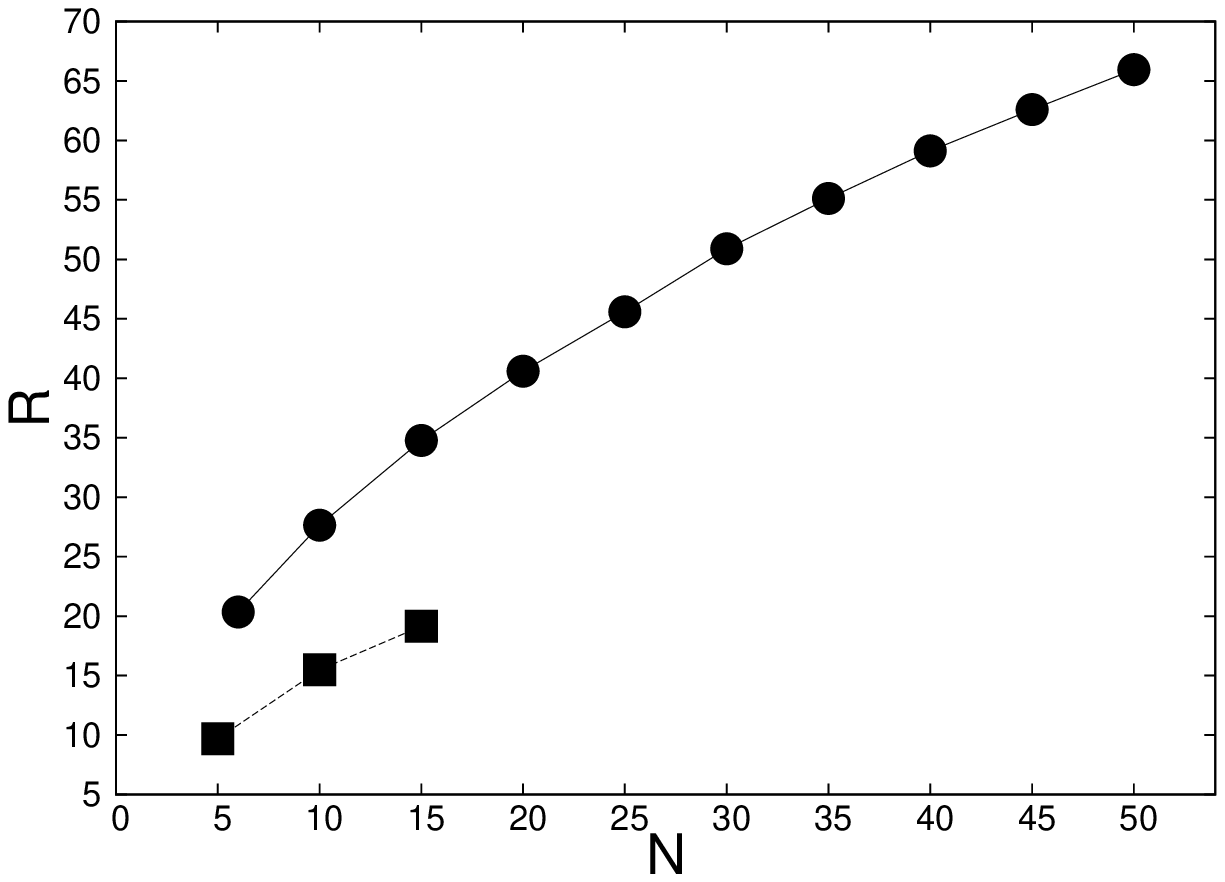}
\caption{The vorton radius $R$ as a function of winding $N,$ 
for $Q=13000$ (circles) and $Q=3000$ (squares).
Note the limited range of values of $N$ for which a vorton solution exists.
}
\label{fig-RN}
\end{center}
\end{figure}

We have been able to construct vorton solutions similar to those presented
above for a range of values of $Q$ and $N.$ 
An illustration of how the vorton properties vary as the physical
quantities are changed is provided by 
Figure~\ref{fig-RN}, which displays the vorton radius $R$ as a function of
$N$ for the two values $Q=13000$ (circles) and $Q=3000$ (squares).
In each case there is a limited range of $N$ for which a 
vorton solution exists, and the width of this range increases with $Q.$
The limited range of $N$ is discussed further in Section~\ref{sec-inf}.

\begin{figure}[ht]
\begin{center}
\includegraphics[width=12cm]{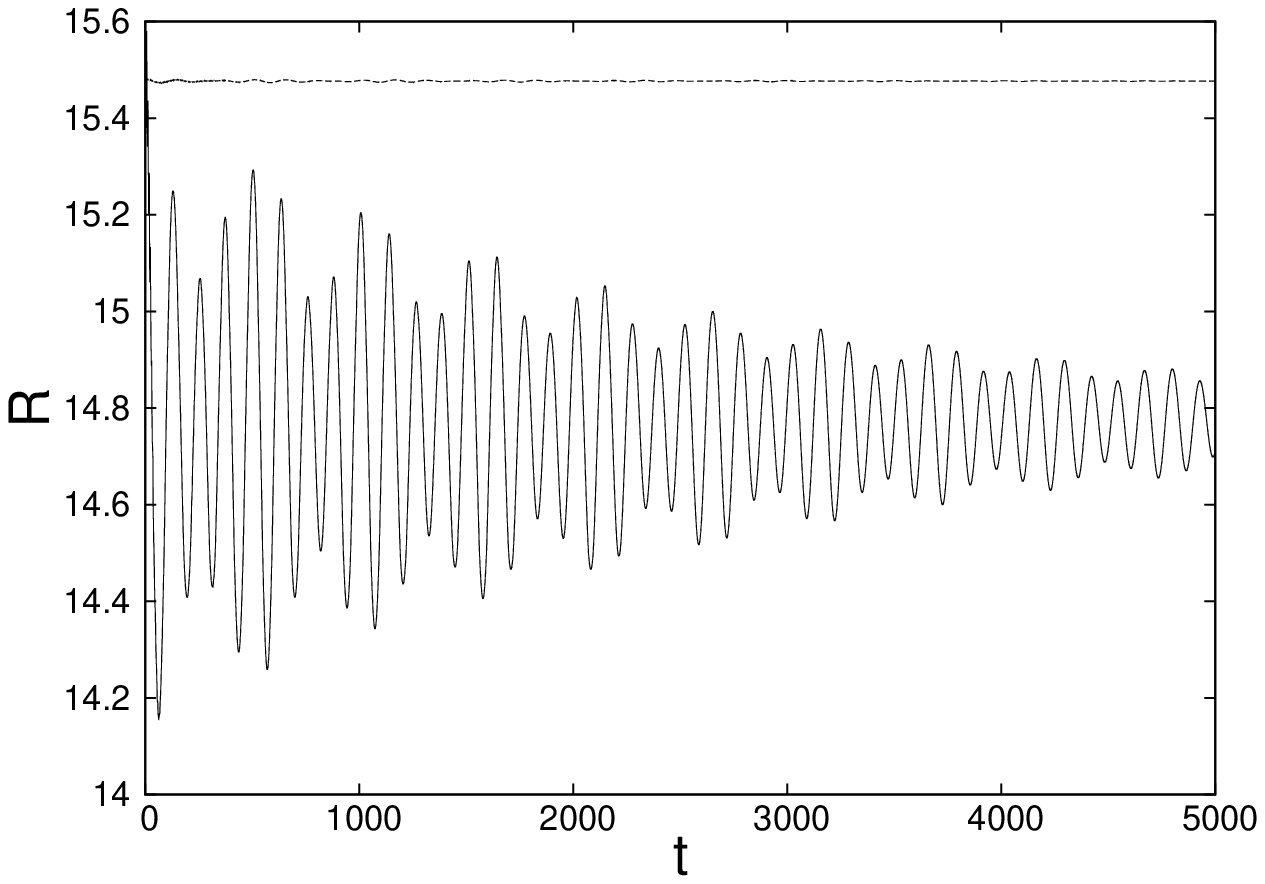}
\caption{The (almost constant) dashed line is the radius $R$ as a 
function of time for the evolution of a vorton with $Q=3000,$ \, $N=10,$ 
and $R=15.5,$ using the correct frequency $\omega=0.88.$ 
The solid curve corresponds to an evolution where the initial
conditions are perturbed by a slight reduction of the initial frequency 
to $\omega=0.83.$
}
\label{fig-RT_2d}
%\includegraphics[width=12cm]{Rt_2d.ps}
%\caption{$N=10,$ $R=27.65$ $\omega=\omega^*=0.68$ and $\omega=0.94\omega^*$}
%\label{fig-f}
\end{center}
\end{figure}

\begin{figure}[ht]
\begin{center}
\includegraphics[width=12cm]{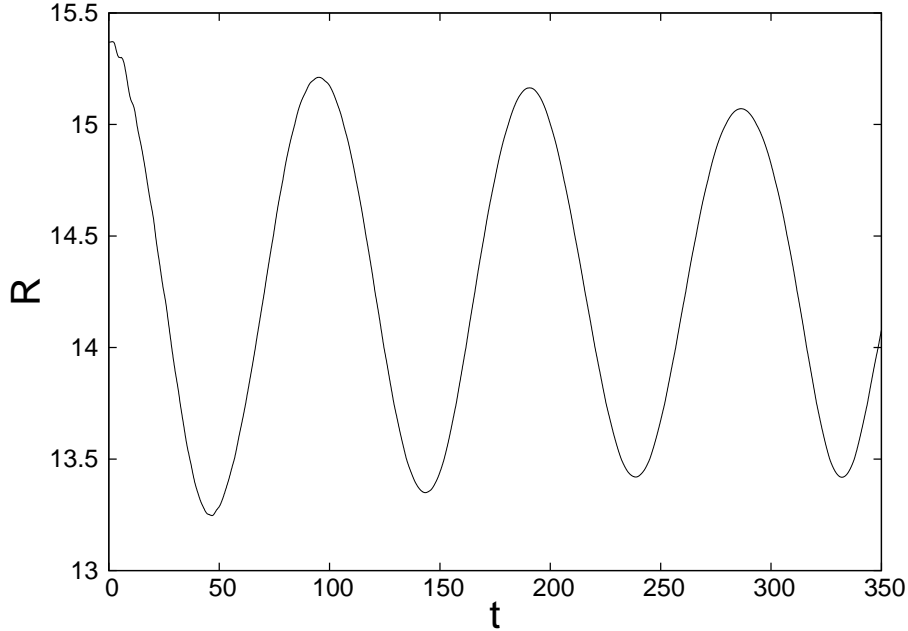}
\caption{The radius $R$ as a function of time in a fully (3+1)-dimensional
simulation of the vorton with $Q=3000$ and $N=10.$
}
\label{fig-RT_3d}
\end{center}
\end{figure}

\begin{figure}[ht]
\begin{center}
\includegraphics[width=15cm]{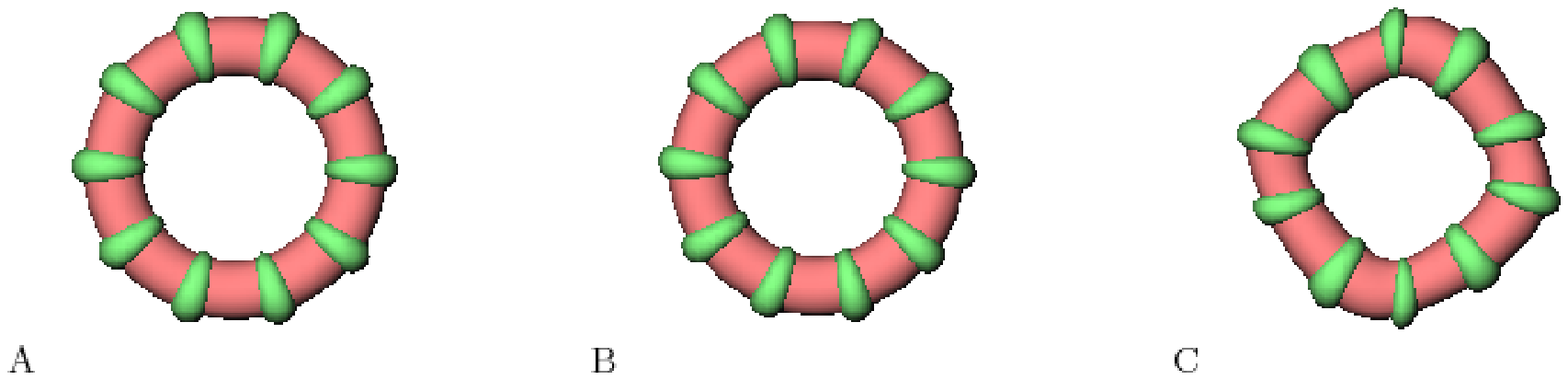}
\caption{The fields at time (A) $t=0$; (B) $t=300$; (C) $t=400$;
for a vorton with $Q=3000$ and $N=10.$ The red (dark) isosurface
is where $|\phi|^2=0.6$ and the green (light) isosurface is 
where $\Re(\sigma)=0.2$}
\label{fig-square}
\end{center}
\end{figure}

\section{Dynamics}\news
A good check of the energy minimization results is to use
the computed fields as initial conditions in a time evolution
simulation and verify that the configuration is indeed stationary.
Restricting to axially symmetric fields allows the \cartoon
method to be used. 

The time evolution is via an explicit second order accurate
finite difference approximation with timestep $\Delta t=0.1.$
At the spatial boundary the fields are held at the vacuum values. 
A small damping term is included, but only on lattice sites which
lie within a small distance ($20\Delta x$) from the edge of the
numerical grid. The role of the damping term is to dissipate any 
radiation that may be generated without slowing the vorton motion.
The frequency $\omega$ is required as part of the initial conditions
but this can be calculated from the energy minimization results
using the relation (\ref{omega}).

As an example, consider the vorton with  $Q=3000,$ \, $N=10,$
and $R=15.5,$ corresponding to the fields presented in Figure~\ref{fig-N10}.
Using the relation (\ref{omega}) the frequency is computed to be
$\omega=\omega_*=0.88.$
Figure~\ref{fig-RT_2d} presents the evolution of the vorton
radius $R$ (almost constant dashed line) as a function of time 
for $0\le t\le 5000.$
This confirms the stationary properties of the initial condition
produced by the energy minimization computation, to a good accuracy.
Similar results have been obtained for other examples.

Next we consider the stability of the constructed vorton solutions
by applying a perturbation to the initial conditions.
At this stage, the evolution algorithm imposes axial symmetry
on the fields so only axially symmetric perturbations can be
investigated using the \cartoon method. Shortly, we shall address
some issues associated with general non-axial dynamics.
 
The solid curve in Figure~\ref{fig-RT_2d} corresponds to an 
evolution where the initial
conditions are perturbed through a slight reduction of the initial 
frequency by $6\%$:  $\omega=0.94\omega_*=0.83.$
Note that even a small perturbation of this kind produces reasonably large
amplitude fluctuations and excites a long-lived
oscillatory mode. There is a slow evolution back to a stationary vorton,
but of course this is not the original unperturbed vorton because
the reduction in the initial frequency means that the initial charge is
reduced from $Q=3000$ to $Q=2820.$ The asymptotic radius that the oscillation
approaches is consistent with the radius of the stationary vorton
with $Q=2820.$
This result is strong evidence
for the stability of this vorton to axially symmetric perturbations,
and similar results have been obtained for other examples.

One of the aims of the current work is to find a vorton with reasonable 
properties that allow a fully (3+1)-dimensional simulation to be performed.
In particular, this is required if stability to non-axial perturbations
is to be investigated.
One condition is that the vorton radius must not be too large in
comparison to its width, otherwise the fine grid resolution required to
accurately approximate the string cross-section would lead to an
unacceptably large number of grid points for a (3+1)-dimensional
simulation to incorporate the whole vorton. In particular, this
means that the thin string limit, which is mainly used in analytical
studies, is beyond current computational feasibility. For (2+1)-dimensional
kinky vortons the thin string limit is computationally accessible \cite{BS}
and the results reveal an excellent agreement with analytical studies.

The vorton presented above with  $Q=3000,$ \, $N=10,$ and $R=15.5$
appears to be a reasonable candidate for a (3+1)-dimensional
simulation. It has a radius which is not much larger than the 
smallest radius vorton we have been able to find, and it seems
a sensible precaution not to choose the smallest possible vorton
as this may be a critical case.
The axially symmetric
energy minimization results can again be used to provide the initial
conditions. 

An additional complication with (3+1)-dimensional simulations
is that the spatial variation of the $\sigma$
field around the loop, which was treated exactly in the axial case
using the ansatz, requires a good resolution. 
To deal with this issue the lattice spacing used in the axial
computations $\Delta x=0.5,$ is reduced to $\Delta x=0.25$ in
the (3+1)-dimensional simulations and, furthermore, the second order 
accurate spatial finite difference approximations are improved to 
fourth order accurate versions. Obviously, the reduction in the
lattice spacing either requires a corresponding increase in the number of grid
points or a reduction in the physical size of the simulation region, or
some combination of the two.

The results of a simulation containing $301^3$ grid points, and
therefore $-L \le x \le L,$ with $L=37.5$, are shown in
Figure~\ref{fig-RT_3d}. Here 
the radius $R$ is plotted as a function of time, for the initially
unperturbed vorton with $Q=3000,$ \, $N=10,$ and $R=15.5$ $(R/L\approx 0.4).$
This Figure reveals that the radius oscillates, which at
first sight may appear contrary to expectations
since the initial conditions were
taken from the energy minimization code. However, the energy minimization
algorithm treats the axial dependence of the fields exactly using the
ansatz (\ref{axial}), and the integration over the polar angle is exact.   
In contrast, the fully (3+1)-dimensional code only treats the axial
dependence approximately and there are discretization errors.
This can be seen by computing the charge $Q$ in the (3+1)-dimensional code,
where it is found to be $Q=2844,$ which is around $95\%$ of 
the true value. One therefore expects to see oscillations, with an
amplitude similar to that seen in the axial simulation of 
Figure~\ref{fig-RT_2d},
where a perturbation of the frequency reduced the charge by a similar amount.
Figure~\ref{fig-RT_3d} shows the expected behaviour of a configuration which
is close to a stationary vorton and slowly approaching the stationary
state.

Figure~\ref{fig-square}A presents the initial fields and
Figure~\ref{fig-square}B is at the later time $t=300$ (after 3000 timesteps).
In these plots the location and thickness of the vorton is displayed 
by plotting the 
red (dark) isosurface $|\phi|^2=0.6$ and the 
winding number of the $\sigma$ field is clear from the green (light) 
isosurface $\Re(\sigma)=0.2$.
For this vorton the period for the internal motion is 
$T=2\pi/\omega=7.14,$ therefore
between the first two plots in Figure~\ref{fig-square} the vorton has made
42 internal revolutions and has retained its original structure.

The vorton disintegrates shortly after the final time displayed in 
Figure~\ref{fig-RT_3d}, due to an instability to non-axial perturbations.
The cubic boundary, and the fact that it reflects back radiation,
provides a small non-axial perturbation that 
eventually has a clear influence on the vorton, as can be seen 
by the square deformation in Figure~\ref{fig-square}C at $t=400.$
As noted above, the boundary is at a distance $L=37.5$ and this is 
not substantially larger than the vorton radius $R=15.5,$ particularly
given the width of the condensate as seen in Figure~\ref{fig-N10}.
The non-axial perturbation produces a pinching and bending of the vorton
which results in its destruction. 

There are several pieces of evidence that support the explanation that
the boundary is providing a symmetry breaking perturbation but
is not responsible for the instability mechanism, which is physical
and not a numerical artifact.
One piece of evidence is provided by repeating
the above simulation with the same lattice spacing $\Delta x=0.25$
but increasing the number of grid points to $501^3$, so that
the boundary is further away at $L=62.5$ $(R/L\approx 0.25$). 
This produces a very similar result, and in particular the vorton exists for almost the same amount of time. 

\begin{figure}[ht]
\begin{center}
\includegraphics[width=12cm]{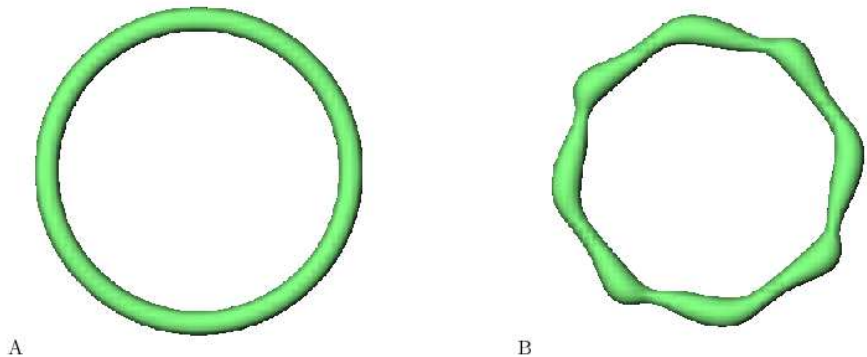}
\caption{The isosurface $|\sigma|^2=0.5$ at time (A) $t=0$; (B) $t=600$;
for a vorton with $Q=13000$ and $N=25.$ Note the pinching 
and bending instability
excited by the cubic boundary.}
\label{fig-conpinch}
\end{center}
\end{figure}

\begin{figure}[ht]
\begin{center}
\includegraphics[width=12cm]{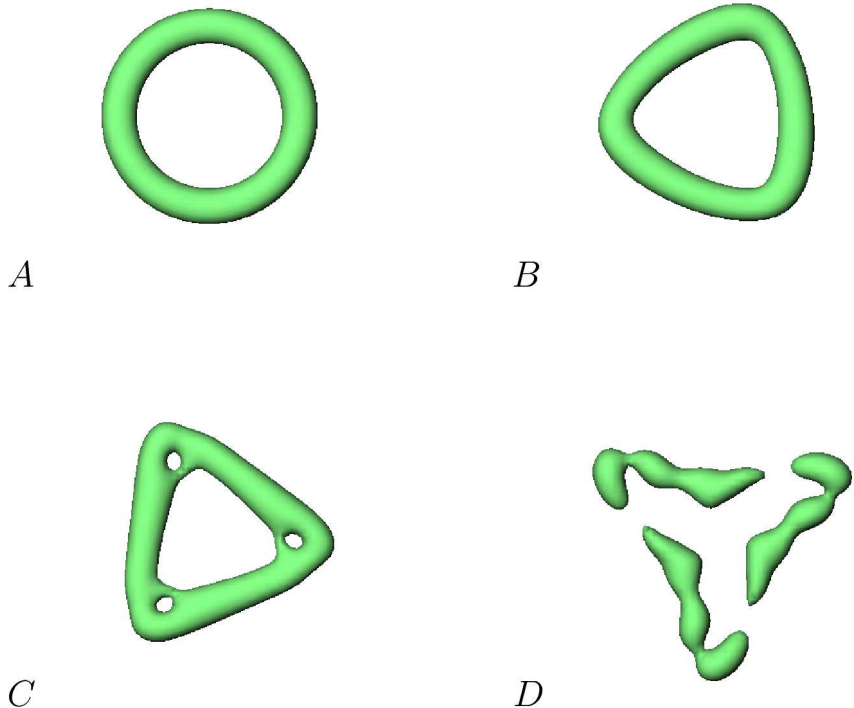}
\caption{The isosurface $|\sigma|^2=0.4$ at time (A) $t=0$; (B) $t=90$;
(C) $t=120$; (D) $t=140$;
for a vorton with $Q=3000$ and $N=10.$ The vorton has an
initial small perturbation with triangular symmetry.}
\label{fig-triangle}
\end{center}
\end{figure}

Further evidence for the instability is obtained by
simulating larger vortons. As an example, consider the vorton
displayed in Figure~\ref{fig-arrows} with $Q=13000$ and 
$N=25,$ giving $R=45.6.$ This has been simulated using the (3+1)-dimensional
code with a lattice spacing $\Delta x=0.5$ and $301^3$ lattice points,
therefore $L=75$ ($R/L\approx 0.6$).
Figure~\ref{fig-conpinch} displays the condensate
isosurface $|\sigma|^2=0.5$ at the initial time $t=0$ and at the
later time $t=600.$  The perturbation, with square symmetry,
induced by the boundary of the grid, is clearly evident together
with the pinching and bending modes. 
Some parts of the vorton cross-section are 
pinched, while others expand to maintain the total constant charge $Q$
carried by the condensate. Shortly after $t=600$ this vorton decays
due to this instability.

Finally, perhaps the most convincing evidence is provided by applying
a small non-axial perturbation, to dominate over the tiny 
square perturbation induced by the boundary of the grid.
Figure~\ref{fig-triangle} displays the isosurface $|\sigma|^2=0.4$ 
at increasing times for a vorton with $Q=3000$ and $N=10.$ The vorton has an
initial small perturbation with triangular symmetry obtained by
scaling the condensate modulus as
\be
|\sigma|\mapsto |\sigma|(1+\varepsilon\sin 3\theta),
\ee
where $\theta$ is the cylindrical polar angle introduced earlier
and the amplitude is $\varepsilon=0.02.$
The perturbation is small and therefore difficult to detect
in the initial condition Figure~\ref{fig-triangle}A. 
However, by  $t=90$ (Figure~\ref{fig-triangle}B)
the triangular deformation has grown to a visible amount,
and by $t=120$ (Figure~\ref{fig-triangle}C) the pinching and bending
instability is apparent. This Figure reveals that the bending 
and pinching combination allows a form of vortex/anti-vortex annihilation
to take place and this ultimately destroys the vorton  by $t=140$
(Figure~\ref{fig-triangle}D). Note that this perturbation, although
small, is easily sufficient to dominate over the tiny square perturbation
provided by the boundary. The vorton disintegrates in a much shorter
time than with the earlier boundary generated perturbation, and the
triangular symmetry is well-preserved, demonstrating that the boundary
has a negligible influence in this case.

Studies of both infinite straight strings \cite{LS}
and (2+1)-dimensional kinky vortons \cite{BS} both demonstrate that 
unstable pinching modes can
exist which require breaking the axial symmetry.
Similar unstable pinching modes are also found in ferromagnetic vortex
rings \cite{Su}, in certain parameter regimes.
Furthermore, unstable non-axial modes can also exist within
elastic string models \cite{CM}. Numerical studies of elastic string
dynamics \cite{MP} display bending instabilities, associated
with curvature growth, and the qualitative features are remarkably
similar to those found in the field theory simulations presented here. 
 It is therefore disappointing, though
not surprising, to find unstable modes that involve breaking the axial 
symmetry. 

It appears that all the vortons constructed in this paper have a
non-axial instability, but this does not rule out the possibility that 
other vortons with different values of $Q$ and $N,$ or 
different parameters of the theory, may be stable.
Unfortunately, as mentioned earlier, most regions are not
accessible numerically with current computational resources,
and indeed it required considerable investigation to find any examples 
of stationary vortons which are stable even to axially symmetric
perturbations. 

Note that the vorton simulations presented in the modified model
\cite{LS} also fail on a time scale that is of a similar order,
but it is difficult to know if the reasons are related.

\section{Infinite string analysis and parameters}\news\label{sec-inf}
One approach to the analysis of vortons is to consider the large
radius limit and approximate the vorton cross-section by
that of an infinite straight string carrying current and charge \cite{LS}.
For an infinite string that lies along the $z$-axis, the appropriate
ansatz in cylindrical polar coordinates is
\be
\phi=e^{i\theta} |\phi|, \quad\quad \sigma=e^{i(\omega t+kz)}|\sigma|,
\label{string}
\ee
where $|\phi|$ and $|\sigma|$ are profile functions that depend
only on $\rho.$ Here $k$ is the rate of twisting and is responsible
for the current along the string.
It is easy to see that the equations that determine the profile
functions depends on $k$ and $\omega$ only through the combination
$\chi=\omega^2-k^2.$
In this approach the key aspect is to determine how properties of
the profile functions, and also various cross-sectional integrals
of these, depend
on $\chi.$ The connection to a vorton loop is then made by 
identifying the twist rate as $k=N/R,$ where $R$ is the radius of the
vorton.

From the form of the ansatz (\ref{string}), it is clear that
in terms of the Lagrangian density (\ref{lag}) the current
and charge together produce a shift by $\chi$ in the effective mass
squared of the $\sigma$ field. Defining the quadratic coefficients
of the two fields by
\be
m_\phi^2\equiv\frac{\lambda_\phi}{2}\eta_\phi^2, \quad
m_\sigma^2\equiv\frac{\lambda_\sigma}{2}\eta_\sigma^2+\chi,
\label{mass}
\ee
then the requirement that the $U(1)$ symmetry of the $\phi$ field
is broken and that of the $\sigma$ field is unbroken yields
the constraint \cite{LS}
\be
\frac{m_\phi^4}{\lambda_\phi}>\frac{m_\sigma^4}{\lambda_\sigma}.
\label{cond1}
\ee
Note that this constraint for infinite straight strings
generalizes the earlier constraint (\ref{convac}), which
is recovered if $\chi=0.$ Other constraints can also be obtained,
by similar arguments, and this leads to an allowed interval
in which $\chi$ must lie.
With the parameters (\ref{param3}) used in this paper, 
the constraint (\ref{cond1}) imposes the upper bound
\be
\chi<\sqrt{\frac{3}{2}}-1=0.22.
\label{bound}
\ee
\begin{figure}[ht]
\begin{center}
\includegraphics[width=12cm]{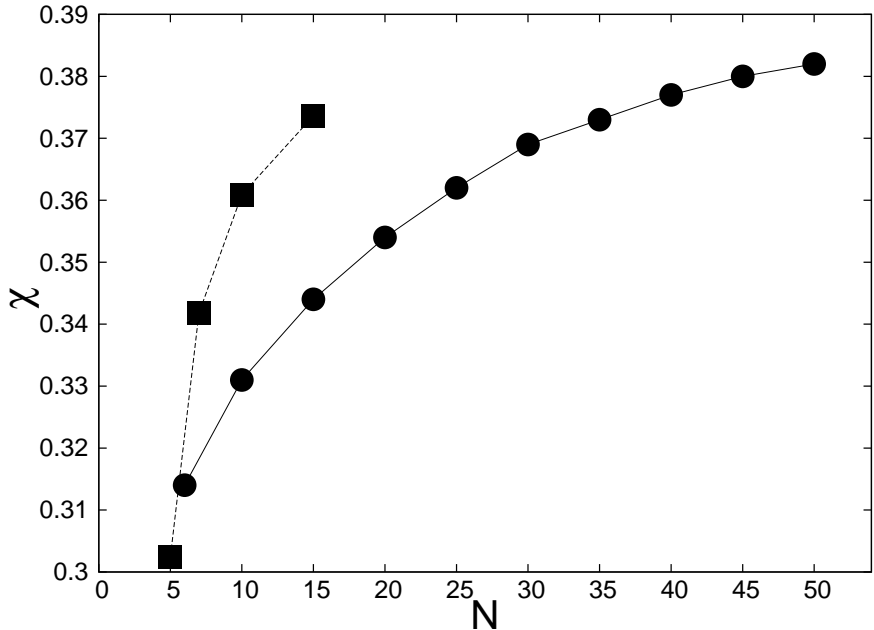}
\caption{$\chi$ as a function of $N$ for vortons with 
 $Q=13000$ (circles) and $Q=3000$ (squares).}
\label{fig-chi}
\end{center}
\end{figure}

Given the numerical results for the vortons constructed using energy 
minimization then $\omega,$ $N$ and $R$ are all known for a given vorton and
therefore $\chi=\omega^2-N^2/R^2$ can be calculated.
The results are displayed in Figure~\ref{fig-chi}, where it can be
seen that all the obtained values of $\chi$ violate the infinite
string bound (\ref{bound}). This is a very surprising revelation
and leads to the conclusion that the cross-section of an infinite 
straight string does not provide a good description of the 
cross-section of the vortons found in this paper.

Note that $\chi$ lies in a very small range, suggesting that there
is an allowed interval, but it is much higher
than that predicted by the straight string analysis. 
The fact that all the vortons found have similar values of $\chi$
also explains the earlier observation that profile functions,
such as those displayed in Figure~\ref{fig-N45} and Figure~\ref{fig-N10},
appear similar after a translation: it is because they have similar
values of $\chi$ and hence similar widths and amplitudes.
It is interesting that for different values of $Q$ the allowed
range for $N$ seems to correspond to the same range for $\chi,$
suggesting that there is the same allowed interval for any $Q.$

We do not have a rigorous explanation of why the straight string analysis
appears not to be applicable to these vortons. However, it seems
likely that it may be related to the fact that in the global theory
a single vortex string has infinite energy per unit length.
Only the combination of a vortex/anti-vortex pair has a finite
energy, therefore calculating properties based on isolating
the vortex and ignoring the anti-vortex is probably not applicable
unlike in the local theory. Although kinky vortons are also
based on a global theory, in that case a single kink string
does have finite energy per unit length, and therefore in this
respect behaves more like the local vorton case.
The straight string analysis does provide an excellent description
for kinky vortons, which supports the idea that the discrepancy
is a result of the divergent energy properties of a global vortex, and we believe it is likely that it will also work for local vortons.
 
Finally, in this section, we briefly mention the results of some
investigations using other parameter values in the Lagrangian.
Consider a one-parameter family of theories given by
 \be \eta_\phi=1, \quad
\eta_\sigma=1+\left(\frac{\sqrt{3}}{2}-1\right)\mu, \quad \lambda_\phi=3-\mu, \quad
\lambda_\sigma=2, \quad \beta=2-\mu \label{param4}. \ee
If $\mu=0$ then we recover the parameter set (\ref{param3})
used throughout this paper, whereas if $\mu=1$ then the parameter set
is that used extensively in the study of kinky vortons \cite{BS}.

Using the energy minimization approach we can attempt to compute
vortons as $\mu$ increases away from zero, using $\mu=0$
vortons as initial conditions, and working piecemeal
towards $\mu=1.$ However, we find that as $\mu$ increases the solution
changes very little until a critical value is reached, which in each case
is less than one, where a solution fails to exist. For example,
for the vorton with $Q=13000$ and $N=10$ the solution exists
up to the critical value $\mu=0.83,$ but for $Q=3000$ and $N=10$
the solution only exists up to $\mu=0.2.$ It, therefore, appears that 
there are regions in parameter space where solutions for given
values of $Q$ and $N$ exist, and this makes the study of vortons
particularly difficult. For example, we have not been able to construct
any vortons with the parameter set given by $\mu=1,$ and this is consistent
with the piecemeal results, though we have no detailed understanding of this.
 In fact this is the point where the phase separation condition 
($\beta^2>\lambda_\phi\lambda_\sigma/4$) discussed in ref.~\cite{BCS} is
 violated and this could be part of the explanation. At this stage it 
appears a very difficult problem to understand and explain the parameter 
values that will and will not allow vortons. 
 
\newpage
\section{Conclusion}\news
We have used energy minimization methods to compute stationary vortons in the
 global version of Witten's $U(1)\times U(1)$ theory and verified that 
they are stable to axially symmetric perturbations using dynamical 
simulations. The study of dynamics without axial symmetry reveals that there are
 unstable modes that are excited by non-axial perturbations. We have also 
found that an analysis based on infinite straight strings does not seem 
to provide a good description of the vortons we have computed.

The cosmological and astrophysical applications of vortons require that at 
least some of the loops formed by the evolution of a cosmic string network 
survive for a substantial fraction of cosmic time. Our results provide some 
progress in clarifying this issue, but they do not unequivocally solve the 
problem. Most importantly, we have found a regime of parameter space where 
vortons with a radius which can be accommodated in a sensible sized numerical 
grid can be constructed. We believe that the reason that the infinite string 
analysis does not yield accurate predictions  for the global version of the 
theory is because of long range interactions associated with a divergence in 
the energy per unit length. The stability analysis we have performed is far 
from exhaustive, although we do believe that the independence of the results 
on $R/L$ suggests that the proximity of the boundary is not the reason for 
the instability. In the case where the perturbations are created by the cubic 
symmetry of the numerical grid, the loop is subjected to a wide range of 
small amplitude modes and it may be that only some of
these modes are unstable. 
Further detailed analysis of the stability properties is clearly needed. 
It could also be that the very narrow range of $\chi$ for which solutions 
can be found plays some role in the apparent sensitivity to perturbations. 
It would, of course, be of considerable interest to generalize the results 
in this paper to Witten's original local theory. In particular, we suspect 
that the failure of the string analysis is due to the properties of a 
global vortex and therefore the string analysis should be successful in the 
local theory.

We note that vorton-like stationary solutions have been constructed in a very 
different regime to those discussed in this paper; and those likely to be 
relevant to cosmology \cite{RV}. In particular, the vorton radius is 
roughly the same as the string width, so the solutions are as far from the 
thin string limit as is possible. Furthermore, only solitons with very 
low winding numbers,
$N\le 5$, could be constructed and it was acknowledged that {\em the
construction of vortons in the thin ring limit remains a numerical
challenge.} It would be interesting to see if a connection can be made between
these two different regimes of vortons, however, given our earlier results
on parameter variations, it seems unlikely that the vortons of one
regime will survive all the way to those of a quite different regime.
The vortons in \cite{RV} were constructed by perturbing
away from the sigma model limit
$\lambda_\phi=\lambda_\sigma\rightarrow\infty$ and
$\beta\rightarrow\infty$ and it was noted that the solutions
are essentially the Skyrmion solutions obtained by the current
authors \cite{BCS} in a related model of a two-component
Bose-Einstein condensate. The solutions constructed in the current
paper are not well-described by the sigma model approximation: 
for example, for the solution with $Q=3000$ and $N=10$ 
it is found that $|\phi|^2+|\sigma|^2$ ranges over the entire
interval $[0.64,1.26],$ and therefore has a substantial variation.

\section*{Acknowledgements}
The parallel computations were performed on 
the Durham University HPC cluster HAMILTON, and
COSMOS at the National Cosmology Supercomputing Centre in Cambridge.

\noindent PMS thanks the STFC for support under the rolling grant ST/G000433/1.\\


\begin{thebibliography}{99}

\bibitem{cartoon} M. Alcubierre, S. Brandt, B. Br\"ugmann, D. Holz,
E. Seidel, R. Takahashi and J. Thornburg, 
\textit{Int.J.Mod.Phys.} \textbf{D10}, 273 (2001). 

\bibitem{BCS}
R.~A. Battye, N.~R. Cooper and P.~M. Sutcliffe, \textit{Phys. Rev.
Lett.} \textbf{88}, 080401 (2002); JHEP, PRHEP unesp2002/009 (2002).

\bibitem{BS}
R.~A. Battye and P.~M. Sutcliffe, 
\textit{Nucl. Phys.} \textbf{B805}, 287 (2008).

\bibitem{BCDT}
R. Brandenburger, B. Carter, A. Davis, M. Trodden,
\textit{Phys. Rev.} \textbf{D54}, 6059 (1996).

\bibitem{CM}
B. Carter and X. Martin,
\textit{Annals Phys.} \textbf{227}, 151 (1993). 

\bibitem{DS2}
R.~L. Davis and E.~P.~S. Shellard,
\textit{Phys. Lett.} \textbf{B209}, 485 (1988).


\bibitem{IS}
T. Ioannidou and  P.~M. Sutcliffe,
\textit{Physica} \textbf{D150}, 120 (2001).


\bibitem{LS}
Y. Lemperiere and E.~P.~S. Shellard,
\textit{Nucl. Phys.} \textbf{B649}, 511 (2003);
\textit{Phys. Rev. Lett.} \textbf{91}, 141601 (2003).


\bibitem{MP}
X. Martin and P. Peter,
\textit{Phys. Rev.} \textbf{D61}, 043510 (2000). 


\bibitem{RV}
E. Radu and M.~S. Volkov, 
\textit{Phys. Rep.} \textbf{468}, 101 (2008).

\bibitem{Su}
P.~M. Sutcliffe, \textit{Phys. Rev.} \textbf{B76}, 184439 (2007).

\bibitem{VS} A. Vilenkin and E.~P.~S. Shellard, \textit{Cosmic Strings
and other Topological Defects}, Cambridge University Press, 1994.

\bibitem{Wi1} E. Witten, \textit{Nucl. Phys.} \textbf{B249}, 557 (1985).

\end{thebibliography}
\end{document}